\documentclass[conference]{IEEEtran}
\IEEEoverridecommandlockouts

\usepackage{cite}
\usepackage{amsmath,amssymb,amsfonts}
\usepackage{color,soul}
\usepackage{graphicx}
\usepackage{physics}
\usepackage{textcomp}
\usepackage{xcolor}
\usepackage{algorithm}
\usepackage{algpseudocode}
\usepackage{xspace}
\usepackage{hyperref}
\usepackage{multirow}
\usepackage{pifont}
\usepackage{tikz}
\usepackage{bm}
\usepackage[table,xcdraw]{}
\newcommand*\circled[1]{%
  \tikz[baseline=(char.base)]{
    \node[shape=circle,fill,inner sep=0pt,minimum size=1em] (char) 
    {\raisebox{0.15ex}{\vphantom{A}\hspace{0.1em}\textcolor{white}{#1}\hspace{0.1em}}};}}
    
\def\BibTeX{{\rm B\kern-.05em{\sc i\kern-.025em b}\kern-.08em
    T\kern-.1667em\lower.7ex\hbox{E}\kern-.125emX}}

\begin{document}

\title{Guess, SWAP, Repeat — Capturing Quantum Snapshots in Classical Memory}

\author{\IEEEauthorblockN{Debarshi Kundu$\bm{^\dagger}$}
\IEEEauthorblockA{\textit{Computer Science and Engineering} \\
\textit{Pennsylvania State University}\\
State College, USA \\
dqk5620@psu.edu}
\and
\IEEEauthorblockN{Avimita Chatterjee$\bm{^\dagger}$}
\IEEEauthorblockA{\textit{Computer Science and Engineering} \\
\textit{Pennsylvania State University}\\
State College, USA\\
amc8313@psu.edu}
\and
\IEEEauthorblockN{Swaroop Ghosh}
\IEEEauthorblockA{\textit{Dept. of EECS} \\
\textit{Pennsylvania State University}\\
State College, USA \\
szg212@psu.edu}
\thanks{$\bm{\dagger}$ Both authors contributed equally to this research.}

}

\maketitle

\begin{abstract}
In this work, we introduce a novel technique that enables the observation of quantum states without directly measuring, and thereby destroying them. Our method enables the observation of multiple quantum states at different points within a single circuit, one at a time, allowing them to be saved into a classical memory without direct measurement or destruction. These states can then be accessed on demand by downstream applications during execution, introducing a dynamic and programmable notion of quantum memory that supports modular, non-destructive quantum workflows. The primary contribution of this work is a hardware-agnostic, machine-learning-driven framework for capturing quantum \texttt{`snapshot'} i.e. non-destructive estimate of quantum state at arbitrary points within a quantum circuit, and enabling their classical storage and later reconstruction, akin to memory operations in classical computing. This capability is critical for real-time introspection, debugging, and memory functionality in quantum systems, yet it remains fundamentally challenging due to the no-cloning theorem and the destructive nature of quantum measurement.
This work introduces a `guess-and-check' methodology that utilizes fidelity estimation via the SWAP test to guide state reconstruction. We introduce both gradient-based deep neural networks and gradient-free evolutionary strategies to estimate quantum states using fidelity alone as the learning signal. 
We implement and validate a key component of our framework on current IBM quantum hardware, achieving high-fidelity ($\sim 1.0$) state reconstructions for Hadamard and other known states. In simulation, our models achieve an average fidelity of $0.999$ over 100 random quantum states. 
These reconstructed quantum states can be stored classically and later reloaded into quantum circuits, providing a realistic path toward long-term, non-volatile quantum memory, establishing a practical and generalizable method for quantum state storage, and laying the foundation for future quantum memory architectures.
\end{abstract}


\section{Introduction}

\begin{figure*}
    \centering
    \includegraphics[width=0.9\linewidth]{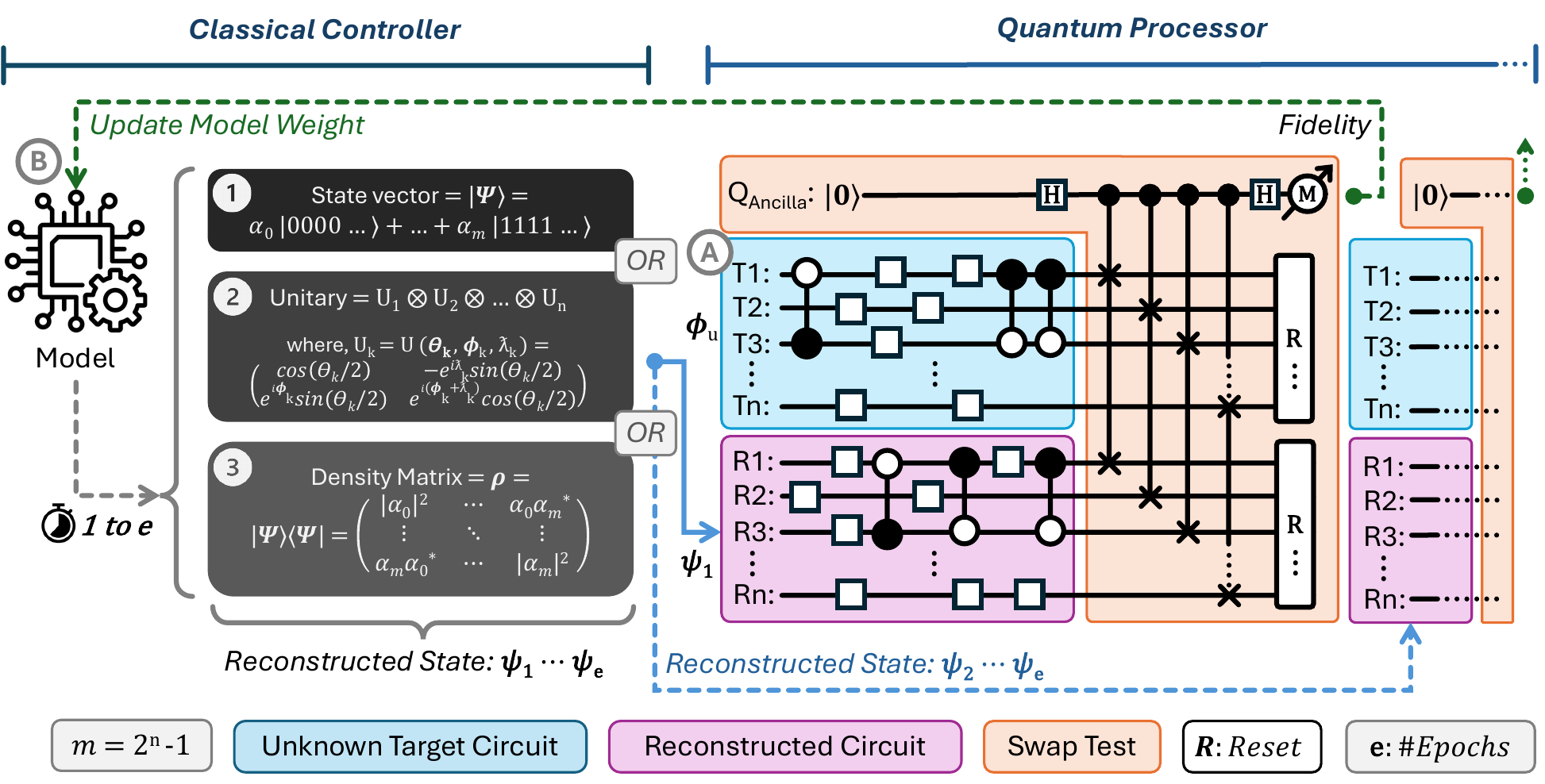}
    \caption{
    \textbf{Methodology for Non-Destructive Quantum State Observation (`Snapshot'):}
    This figure illustrates the methodology behind the three primary processes for observing a quantum state with $n$ qubits. The approach begins with an unknown target circuit (marked in light blue) and a neural network model that generates an output representing either a state vector (Method 1), a unitary (Method 2), or a density matrix (Method 3). This output is used to construct a reconstructed circuit (marked in light purple).  
    An ancilla qubit is initialized in the state \( |0\rangle \), and a swap test is performed between the target and reconstructed circuits. The outcome is projected onto the ancilla qubit, and its measurement yields 
 fidelity, a measure of similarity between the two states. A fidelity of 1 indicates a perfect match between the reconstructed and target states. If the fidelity is not sufficiently close to 1, the neural network’s model weights are updated using the fidelity value. This iterative process continues until the fidelity converges to a value very close to 1, ensuring accurate state reconstruction.
}
    \label{fig:process}
\end{figure*}

Capturing a quantum snapshot, i.e., the act of identifying a quantum state without destroying it, is currently beyond the reach of practical quantum technology. Likewise, there is no reliable method to implement quantum memory to store an arbitrary quantum state indefinitely. Yet, these capabilities are desirable for the maturation of quantum technologies. They promise transformative potential in areas such as quantum communication \cite{heshami2016quantum}, control\cite{riste2012feedback}, debugging\cite{erhard2019characterizing}, and real-time circuit introspection\cite{hebenstreit2015real}.

At the heart of quantum computation lies a fundamental challenge: the ability to observe and understand the state of a quantum system. Unlike classical systems, where observation is passive and non-disruptive, quantum systems exhibit a profound fragility. The act of measurement does not merely reveal information, it irreversibly alters the very state being observed. This phenomenon, known as wavefunction collapse, ensures that once a quantum state is measured, its original form ceases to exist. As a result, direct measurements both destroy the intricate superpositions and entanglements that underpin quantum advantage and yield only partial information about the system.

\textbf{Motivation:}
A close neighbor of the quantum snapshot concept is Quantum State Tomography (QST). Traditionally, QST has been the principal technique for reconstructing a complete description of a quantum state, typically as a density matrix, by performing repeated measurements on many identically prepared copies of the system. It plays a critical role in verifying prepared states, benchmarking quantum hardware, and characterizing noise in quantum processes, and has become a standard tool for experimental validation. However, QST remains fundamentally destructive: it relies on projective measurements that collapse the quantum state during the process of reconstruction. Consequently, it cannot serve as a mechanism for capturing quantum snapshots, where the goal is to capture the state non-destructively and without requiring multiple copies. Besides, QST is a painstakingly slow and tedious process. Full quantum state tomography (QST) typically requires \( 3^n \) distinct measurement settings—one for each combination of Pauli basis on \( n \) qubits—which implies the need for that many copies of the same quantum state~\cite{smith_efficient_2021}. 
This calls for new methods of quantum state observation that preserve coherence with evolving quantum information \cite{von_lupke_parity_2022}.

Realizing quantum memory (a related concept to QST and quantum snapshot) is another major bottleneck in quantum computing, as current qubit technologies suffer from short coherence times and quantum measurements irreversibly collapse state superpositions~\cite{schlosshauer2019quantum, gundougan2023ultimate}. Practical quantum memories, though highly desirable for applications such as quantum repeaters, distributed quantum computing, and quantum internet~\cite{kimble2008quantum, sangouard2011quantum, wehner2018quantum}, remain technologically elusive. Given these limitations, a natural question arises: \textit{Can quantum information be extracted and stored classically, without collapsing the state, such that quantum applications can access it later during computation?}

\textbf{Contribution:}
In this work, we propose a method for taking quantum snapshots to enable non-simulatneous dynamic observation and reconstruction of quantum states during circuit execution. Crucially, this is achieved without collapsing or destroying the quantum state, allowing multiple states at multiple points within the circuit to be accessed and stored. Our method estimates the quantum state and saves it into classical memory. These stored states can later be retrieved or re-prepared by downstream applications as needed.
The primary focus of this work is the capturing of \textit{quantum `snapshots' i.e., estimation of a quantum state at specific points during circuit execution}, and storing them in classical memory. Our approach is designed to be compatible with current quantum hardware architectures~\cite{IBMQuantumCentric2024}. 
%
This approach has broader implications, particularly for the classical storage and reuse of quantum information. By reconstructing and storing quantum states as classical vectors, our method lays the foundation for a form of classical quantum memory. These reconstructed states can be reloaded by quantum applications requiring access to previously encountered states, such as those employing standard Quantum RAM (QRAM) \cite{giovannetti_quantum_2008, phalak_quantum_2023} retrieval algorithms, thus enabling non-destructive reuse without relying on persistent physical quantum memory.

\begin{table}
\fontsize{7.5pt}{9.0pt}\selectfont
\centering
\caption{Mapping Quantum Representations to Estimation Methods}
\begin{tabular}{cc}
\hline
\multicolumn{2}{|c|}{\textbf{Quantum State Representation}}                                                                                                                                                                               \\ \hline
\multicolumn{2}{|c|}{(i) State Vector (Pure)}                                                                                                                                                                                             \\
\multicolumn{2}{|c|}{(ii) Density Matrix (Pure + Mixed)}                                                                                                                                                                                  \\
\multicolumn{2}{|c|}{(iii) Unitary Matrix (From $\ket{0}^{\otimes n} \rightarrow \ket{\psi}$)}                                                                                                                                            \\ \hline
\bm{$\downarrow$}                                                                                                             & \bm{$\downarrow$}                                                                                                   \\
\multicolumn{2}{c}{\textbf{Each representation is reconstructed using:}}                                                                                                                                                                  \\
\bm{$\downarrow$}                                                                                                             & \bm{$\downarrow$}                                                                                                   \\ \hline
\multicolumn{2}{|c|}{\textbf{Estimation Strategies}}                                                                                                                                                                                      \\ \hline
\multicolumn{1}{|c|}{\textbf{(a) Gradient-Based}}                                                                        & \multicolumn{1}{c|}{\textbf{(b) Gradient-Free}}                                                                \\ \hline
\multicolumn{1}{|c|}{\begin{tabular}[c]{@{}c@{}}- Deep Learning Generator \\ - Trained via backpropagation\end{tabular}} & \multicolumn{1}{c|}{\begin{tabular}[c]{@{}c@{}}- Evolutionary Strategies \\ - Optimization-based\end{tabular}} \\ \hline
\end{tabular}
\label{tab:estimation_with_representation}
\end{table}

The primary contributions of this work are:
%
    \circled{1}
    \textit{Non-destructive Quantum Snapshot:} 
We accomplish this through a controlled feedback loop, where a machine learning model iteratively generates candidate quantum states and receives guidance based solely on fidelity scores computed via the SWAP test. This fidelity acts as a non-destructive signal indicating how close the generated state is to the unknown target state. The model updates its parameters based on this feedback, gradually improving its reconstruction accuracy. This controlled, iterative process enables quantum states to be reconstructed without measurement-induced collapse, thus laying the foundation for real-time, introspective quantum computation.
\circled{2}
    \textit{Proof of Concept on Real Quantum Hardware:} 
    We implement and validate a core component of our approach i.e., state reconstruction from fidelity measurements, on IBM superconducting qubit hardware. We demonstrate practical reconstruction using repeated state preparation, achieving fidelities exceeding 99.9\% in simulation and up to 99\% on hardware.
\circled{3}
    \textit{Learning Quantum States Using Fidelity as the Only Signal:} 
    We develop and evaluate two complementary machine learning strategies: \textbf{(i)} a gradient-based deep neural network, and \textbf{(ii)} a gradient-free evolutionary strategy (QESwap), to reconstruct unknown quantum states using only fidelity (measured via the SWAP test) as feedback. Both methods are designed to optimize the state estimate purely from measurement statistics, without needing prior knowledge of the state or its density matrix. Our overall approach is summarized in Table~\ref{tab:estimation_with_representation}, which illustrates the quantum state reconstruction framework using different representations and estimation strategies.
\circled{4}
    \textit{Hardware-aware and Scalable Design for NISQ Devices:} 
    Unlike prior works that rely on access to the full-density matrix of unknown states (impractical) or use synthetic datasets for evaluation, we operate directly on noisy quantum hardware and require no training datasets. This makes our technique well-suited for practical deployment on NISQ-era devices, ensuring compatibility with real-world quantum noise and constraints.
\circled{5}
    \textit{Classical Storage and Retrieval of Quantum States:} 
    We classically store reconstructed quantum states (state vectors) and later reinitialize them on quantum hardware. This mechanism offers a realistic solution to quantum memory limitations for small-scale systems, enabling long-term, non-volatile storage and reuse of quantum information, an essential step toward scalable quantum memory banks and QRAM systems.



\textbf{Paper Structure:}
%
The remainder of this paper is organized as follows. Section II presents related works, highlighting key differences between our approach and existing quantum state reconstruction methods. Section III formally defines the problem and outlines the design goals of our quantum snapshot framework. Section IV describes our methodology, detailing both gradient-based and gradient-free state estimation strategies, and the underlying quantum representations. Section V covers implementation and evaluation across simulated and real quantum hardware, including fidelity benchmarks, entanglement analysis, and convergence performance. Section VI explores practical applications of quantum snapshots, ranging from QRAM to quantum debugging. Section VII discusses limitations, particularly in handling mixed states, and outlines directions for future improvements. Finally, Section VIII concludes the paper by summarizing our contributions and their significance for future quantum memory architectures.

\section{Related Works: A Comparative Perspective}

\begin{table*}[t!]
\centering
\fontsize{8.0pt}{9.5pt}\selectfont
\caption{Comparative Evaluation of the Quantum Snapshot approach and Prior Quantum State Reconstruction Methods}
\begin{tabular}{cl||c|ccccc}
\multicolumn{2}{c||}{\textbf{Metric}}                                                                                                            & \textbf{Snapshot} & \textbf{Liu, et al~\cite{liu_variational_2020}} & \textbf{Ahmed, et al~\cite{ahmed_quantum_2021}} & \textbf{Quek, et al~\cite{quek_adaptive_2018}} & \textbf{Luu, et al~\cite{luu_universal_2024}} & \textbf{Lange, et al~\cite{lange_adaptive_2023}} \\ \hline \hline
\multicolumn{2}{c||}{\textbf{State Preservation}}                                                                                                & \ding{51}            & \ding{55}                           & \ding{55}                             & \ding{55}                            & \ding{55}                           & \ding{55}                             \\ \hline
\multicolumn{2}{c||}{\textbf{Hardware Agnostic}}                                                                                                 & \ding{51}            & \ding{51}                           & \ding{55}                             & \ding{51}                            & \ding{55}                           & \ding{55}                             \\ \hline
\multicolumn{2}{c||}{\textbf{Dataset Requirement}}                                                                                               & \ding{55}            & \ding{55}                           & \ding{51}                             & \ding{51}                            & \ding{51}                           & \ding{51}                             \\ \hline
\multicolumn{1}{c|}{\multirow{2}{*}{\textbf{\begin{tabular}[c]{@{}c@{}}Noiseless\\ Simulation\end{tabular}}}} & \textbf{Min. Fidelity Achieved} & \bm{$0.999$}        & $0.991$                       & $0.9$                           & $0.999$                        & $0.97$                        & $0.956$                         \\ \cline{2-8} 
\multicolumn{1}{c|}{}                                                                                         & \textbf{\# Training Epochs}              & \bm{$5$}           & $240$                         & $1000$                          & \texttt{NA}                           & $1300$                        & $1000$                          \\ \hline
\multicolumn{1}{c|}{\multirow{2}{*}{\textbf{\begin{tabular}[c]{@{}c@{}}Noisy\\ Simulation\end{tabular}}}}     & \textbf{Min. Fidelity Achieved} & \bm{$0.996$}         & \ding{55}                        & \ding{55}                             & \ding{55}                            & $0.3$                         & $0.9093$                        \\ \cline{2-8} 
\multicolumn{1}{c|}{}                                                                                         & \textbf{\# Training Epochs}              & \bm{$5$}          & \ding{55}                         & \ding{55}                             & \ding{55}                            & $1300$                        & $1000$                          \\ \hline
\multicolumn{2}{c||}{\textbf{Quantum Hardware Result}}                                                                                                   & \ding{51}            & \ding{55}                           & \ding{55}                             & \ding{55}                            & \ding{55}                           & \ding{55}                             \\ \hline \hline
\end{tabular}
\label{tab:background_comparison}
\end{table*}

Recent years have seen the emergence of deep learning-based approaches to quantum state reconstruction and tomography methods~\cite{liu_variational_2020, ahmed_quantum_2021, gaikwad_gradient-descent_2025, quek_adaptive_2018, luu_universal_2024, lange_adaptive_2023}. However, these methods exhibit significant shortcomings. 
Table~\ref{tab:background_comparison} compares the proposed approach against existing quantum state reconstruction and tomography methods. Our work is evaluated alongside prominent approaches by Liu et al.~\cite{liu_variational_2020}, Ahmed et al.~\cite{ahmed_quantum_2021}, Quek et al.~\cite{quek_adaptive_2018}, Luu et al.~\cite{luu_universal_2024}, and Lange et al.~\cite{lange_adaptive_2023}. Key metrics include state preservation, hardware agnosticism, dataset requirements, fidelity achieved under noiseless and noisy simulations, number of training epochs, and availability of hardware results. Our approach uniquely supports non-destructive state observation (state preservation), requires no training dataset, and achieves high fidelity with fewer epochs, even on real hardware, highlighting its practical efficiency and generalizability. While the table summarizes key methods in machine learning-based quantum state reconstruction, one approach ~\cite{gaikwad_gradient-descent_2025} has been excluded due to a fundamental incompatibility with the comparison criteria. Gaikwad et al.~\cite{gaikwad_gradient-descent_2025} employ the Uhlmann–Jozsa (UJ) fidelity as a loss function, under the assumption that the density matrix of the unknown quantum state is accessible. However, such access is fundamentally unavailable; indeed, the very motivation behind QST lies in the absence of this information. Consequently, using UJ fidelity in this context is not only impractical but infeasible to compute or implement on actual quantum hardware (and hence excluded from the comparative study).
The comparative analysis in Table~\ref{tab:background_comparison} highlights several key distinctions between our work and existing machine learning-based quantum state reconstruction methods. First and foremost, we uniquely support state preservation, enabling observation without collapsing the quantum state, an essential requirement for dynamic quantum computation. In contrast, all other methods in the comparison table rely on measurement-based approaches that inherently destroy the state, making them incompatible adaptive workflows. Additionally, our approach is hardware agnostic, allowing seamless deployment across different quantum platforms. While some prior methods share this trait, others are constrained by specific hardware assumptions or simulation environments, for example, Ahmed et al.~\cite{ahmed_quantum_2021} present results exclusively on an optical system, limiting the generalizability of their approach. A significant advantage of our approach is its independence from pre-collected datasets; it estimates quantum states in real-time using a feedback-driven process rather than relying on large volumes of training data. This stands in contrast to all compared methods, which require curated datasets, typically from repeated measurements of identically prepared states, to learn state representations.
In terms of noiseless simulation fidelity, we achieve a high minimum fidelity of 0.999 in just 5 epochs, significantly outperforming several methods that require hundreds or even thousands of epochs to reach lower fidelity benchmarks. 
The trend continues under noisy simulation conditions, where we again achieve 0.99 fidelity, whereas the other methods either fail to report results or show considerable degradation in performance. Notably, hardware results are reported only for our approach, demonstrating its real-world feasibility, while all other works remain invalidated on actual quantum devices. Taken together, these comparisons emphasize that our approach not only offers theoretical advantages but also demonstrates superior empirical performance, making it a practical and scalable solution for non-destructive quantum state reconstruction.

\section{Methods}\label{methods}

\subsection{The Pivotal Design Principle}

\begin{figure}
    \centering
    \includegraphics[width=0.9\linewidth]{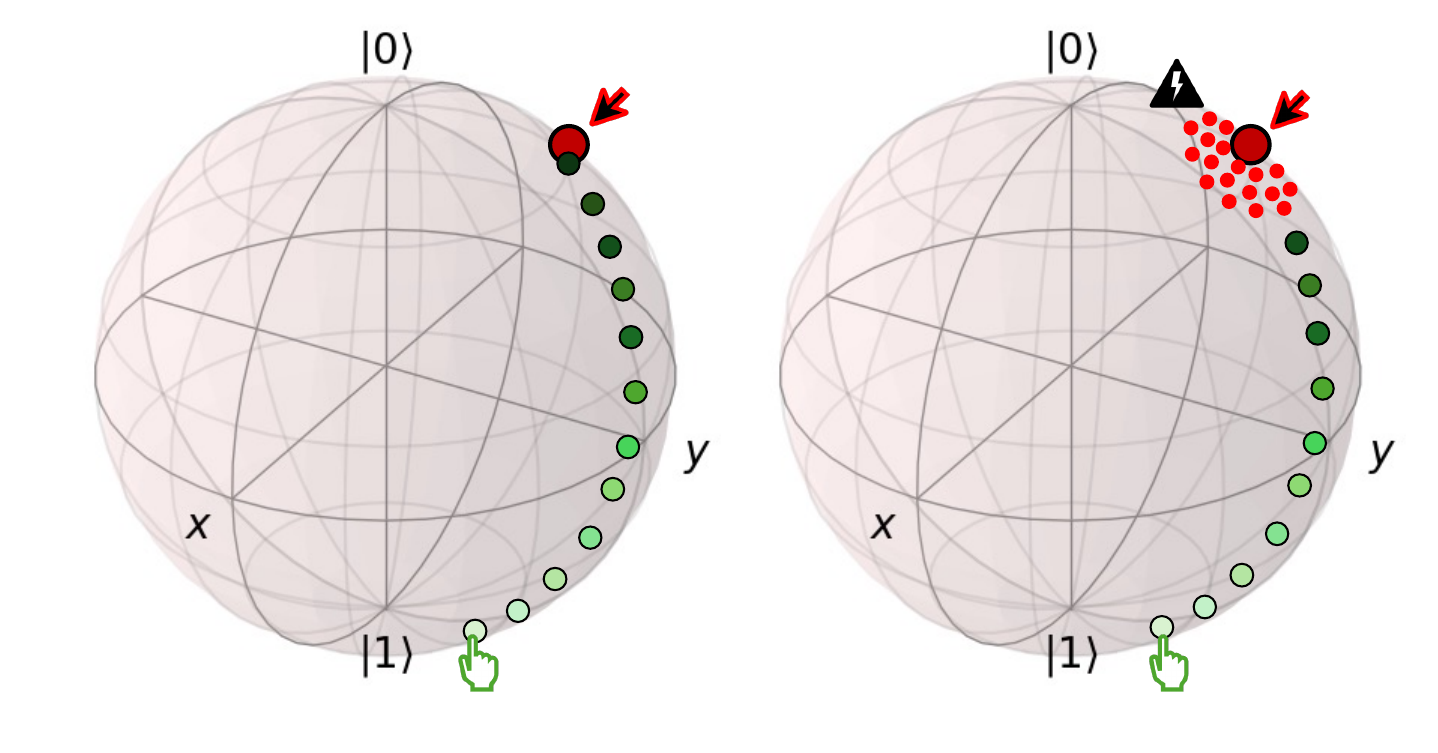}
    \caption{\textbf{Progression of reconstructed states towards target states for noiseless and noisy conditions:} The left figure shows the noiseless case where the target unknown state remains fixed, allowing smooth convergence. In contrast, the right figure illustrates the noisy case, where the target state keeps shifting due to hardware noise, making convergence significantly more challenging.}
    \label{fig:bloch_sphere}
\end{figure}

The unknown quantum state is assumed to be available in
the form of a quantum circuit which is the way of representing the quantum states in modern quantum computers. For the purpose of demonstration, we generate a random quantum state randomly sampled from the complex unit sphere in $\mathbb{C}^{2^n}$ and then prepare the circuit via the Mottonen state preparation method~\cite{mottonen_transformation_2004}, implemented using \texttt{qml.StatePrep}~\cite{pennylane} primitive provided by PennyLane, as illustrated in Figure~\ref{fig:process}A. Subsequently, we initialize a randomly guessed quantum state, which is produced by our classical ML model (Figure~\ref{fig:process}B). This generated state is then loaded into the circuit via \texttt{StatePrep}, alongside an ancilla qubit, which is used to record the result of the SWAP test.

We repeatedly reinitialize the unknown quantum state, thereby requiring multiple copies, an assumption aligned with the principles of quantum state tomography. In each iteration, both the unknown and generated quantum states are freshly initialized. The generated state is derived from the output of the classical model, and a SWAP test is conducted to evaluate fidelity. All qubits, including those encoding the unknown state, are reset and reinitialized for every fidelity measurement. As shown in Table~\ref{tab:estimation_with_representation}, there are $3 \times 2 = 6$ distinct estimation strategies, corresponding to the three types of quantum state representations, state vector; density matrix; and unitary matrix, each of which can be reconstructed using either gradient-based or gradient-free methods. The overall procedure is depicted in Fig.~\ref{fig:process}.


In the case of the \textit{state vector}, we directly estimate $\ket{\psi} \in \mathbb{C}^{2^n}$. For the \textit{density matrix} and \textit{unitary matrix}, we estimate complex-valued matrices of size $2^n \times 2^n$, subject to appropriate constraints (e.g., Hermiticity and trace-one for density matrices, unitarity for unitaries). All reconstruction methods exhibit similar progression on a Bloch sphere representation. An example is shown in Fig.~\ref{fig:bloch_sphere} (left Bloch sphere). The progression begins from a random initial state and gradually converges toward the target state. This illustration reflects the ideal (noiseless) scenario, where the unknown target state remains fixed, enabling smooth and stable convergence.

We first detail the gradient-based approach for estimating the state vector, followed by its gradient-free counterpart. The same estimation framework is then extended to density matrix and unitary generation, with adaptations to accommodate the higher-dimensional structure and relevant physical constraints of each representation.
We describe the procedure for generating the state vector in detail. This approach naturally extends to the other two cases: unitary and density matrix representations. While we provide the overall process in pseudocode, detailed explanations are omitted due to space constraints.

\subsection{Gradient Based Method: Deep Learning Model as Generator}

\begin{algorithm}
\fontsize{8.0pt}{9.0pt}\selectfont
\caption{Gradient-based State Vector Estimation}
\begin{algorithmic}[1]
\Require Number of qubits $N$, epochs $E$
\State $d \gets 2^N$
\State Initialize neural network $\mathcal{F}_\theta: \mathbb{R} \to \mathbb{C}^d$
\State Sample target state $\ket{\psi} \in \mathbb{C}^d$ and normalize

\For{epoch $= 1$ to $E$}
    \State $z \gets \mathcal{U}(0, 1)^{256}$
    \State $\ket{\psi_\theta} \gets \mathcal{F}_\theta(z)$
    \State Use SWAP test to estimate fidelity $F = |\bra{\psi_\theta}\ket{\phi_u}|^2$
    \State Compute loss: $\mathcal{L} = 1 - F$
    \State Update $\theta$ using gradient descent
    \If{$F = \sim 1.0$}
        \State \textbf{break}
    \EndIf
\EndFor

\State \Return final fidelity $F$
\end{algorithmic}
\label{alg:state_vector}
\end{algorithm}

\subsubsection{\textbf{Problem Formulation}}

The goal of this method is to reconstruct an unknown quantum state vector 
$\ket{\psi} \in \mathbb{C}^{2^n}$ given no prior information about its structure. Specifically, we aim to train a model that can generate a quantum state $\ket{\phi}$ such that the fidelity 
\(F = |\bra{\psi}\ket{\phi_u}|^2\)
approaches unity. This indicates that the generated state is indistinguishable from the unknown state (global phase difference doesn't matter).
We develop a deep neural network model where the generator generates the states intending to match with the unknown target state and the SWAP test acts like a loss function telling how far the generated state is similar to the target unknown state thus guiding the generator to train towards generating correct state which brings fidelity between quantum states 1. Algorithm~\ref{alg:state_vector} shows the algorithm for state vector-based fidelity estimation.

\subsubsection{\textbf{Overview of the Method}}
    \textbf{(a)} \textit{Unknown state preparation:} A normalized, randomly sampled quantum state $\ket{\phi_u}$ is generated, along with its corresponding quantum circuit constructed via the Mottonen state preparation method~\cite{mottonen_transformation_2004}, implemented using \texttt{qml.StatePrep}~\cite{pennylane}. This acts as a proxy for an unknown quantum state encountered in practical quantum computations. This step is repeated each time a newly generated candidate state is evaluated, as described in the next step. 
    \textbf{(b)} \textit{Classical neural network generator:} A classical neural network is trained to output a candidate quantum state vector $\ket{\psi}$. This output is normalized and converted into a quantum circuit using the Mottonen state preparation method~\cite{mottonen_transformation_2004}, implemented via \texttt{qml.StatePrep}~\cite{pennylane}.
    \textbf{(c)} \textit{Fidelity estimation via SWAP test:} The overlap between $\ket{\psi}$ and $\ket{\phi_u}$ is estimated using a quantum SWAP test. This test is performed by conducting the SWAP test between a newly generated candidate state and a re-prepared instance of the unknown target state $\ket{\phi_u}$.
    \textbf{(d)} \textit{Loss function and training:} The fidelity computed from the SWAP test defines a loss function that guides the neural network to generate candidate states with maximal overlap with the unknown target. The model is trained using gradient-based optimization, with finite-difference gradient estimation applied due to the non-differentiable nature of the quantum components.
    \textbf{(e)} \textit{Classical storage and retrieval:} Once a candidate state $\ket{\psi}$ achieves high fidelity with the unknown target state $\ket{\phi_u}$, it is stored as a classical vector. This representation can later be reloaded and re-prepared on quantum hardware using standard QRAM-compatible techniques. 

\subsubsection{\textbf{Unknown Quantum State Generation}}\label{unknown_state_prep}
We simulate an arbitrary, unknown quantum state $\ket{\phi_u}$ over $n$ qubits. The state vector is randomly sampled from the complex unit sphere in $\mathbb{C}^{2^n}$. The real and imaginary components of each amplitude are independently sampled from a standard normal distribution. The state vector is then normalized to unit length:
\[
\ket{\phi_u} = \frac{\vec{r} + i \vec{i}}{\|\vec{r} + i \vec{i}\|}
\]
where $\vec{r}, \vec{i} \in \mathbb{R}^{2^n}$ are drawn from $\mathcal{N}(0,1)$.

\subsubsection{\textbf{Generator Architecture}}

The quantum state generator is implemented as a deep fully connected neural network, \texttt{QuantumStateGenerator}, using PyTorch. It takes as input a 256-dimensional latent vector \( \mathbf{z} \) and outputs a real-valued vector of dimension \( 2 \times \texttt{dim} \), where \texttt{dim} denotes the number of complex amplitudes in the target quantum state. The output vector is interpreted as the concatenation of the real and imaginary components of the generated quantum state.
The architecture comprises six linear layers interleaved with GELU (Gaussian Error Linear Unit) activations, except at the output. 

\subsubsection{\textbf{Fidelity-Based Optimization via Swap Test}}\label{swap_test}

Given two quantum states \( \ket{\psi} \) and \( \ket{\phi_u} \), the fidelity based on the quantum \textit{swap test} is defined as:
\(
F(\psi, \phi_u) = |\langle \psi | \phi_u \rangle|^2.
\)
Let \( P(0) \) denote the probability of measuring the ancilla qubit in the state \( \ket{0} \) at the end of the swap test. The fidelity can be expressed in terms of \( P(0) \) as:
\(
|\langle \psi | \phi_u \rangle|^2 = 2P(0) - 1.
\)
Equivalently, let \( \langle Z \rangle \) denote the expectation value of the Pauli-\(Z\) operator on the ancilla qubit. Since
\(
\langle Z \rangle = P(0) - P(1) = 2P(0) - 1
\),
the fidelity can also be written as:
\(
|\langle \psi | \phi_u \rangle|^2 = \langle Z \rangle,
\)
which serves as the primary optimization objective during training.
The swap test employs \( 2n + 1 \) qubits, as illustrated in Fig.~\ref{fig:process}. Qubits \( 1 \) to \( n \) are initialized with the state \( \ket{\phi_u} \), qubits \( n+1 \) to \( 2n \) with \( \ket{\psi} \), and qubit \( 0 \) serves as the control qubit for the controlled-swap operations.
The expected value of the Pauli-Z observable on the control qubit gives:
\begin{equation}\label{eq:fidelity}
\langle Z \rangle = |\bra{\psi}\ket{\phi_u}|^2 \quad \Rightarrow \quad F = \langle Z \rangle
\end{equation}
Since the standard swap test is non-differentiable, we implement a custom autograd function, \texttt{SwapTestFunction}, which uses finite-difference approximation to compute gradients. In the forward pass, the real and imaginary components of the generated output are combined to form a complex-valued state vector, which is normalized to unit norm. The fidelity is then computed between this state and the unknown target.
During the backward pass, each element of the generated output is perturbed slightly in both directions to estimate the gradient via symmetric finite differences. This approach allows us to propagate gradients through the fidelity computation and train the network using standard gradient-based optimization methods.

\subsubsection{\textbf{Combined Model}}
The final stage of the model integrates the generator layer and fidelity evaluation layer. It receives a latent vector \( \mathbf{z} \) and an unknown target state as input and returns both the fidelity score and the normalized generated quantum state. The model is trained to maximize the fidelity, while the generated state is preserved for downstream use in quantum circuit construction.

\subsubsection{\textbf{Training Procedure}}
We train the model using the Adam optimizer with a learning rate of \( 1 \times 10^{-4} \). The loss function is defined as:
\(
\mathcal{L} = 1 - F,
\)
where \( F \) is the fidelity between the generated and target states.
At each training step, a latent vector \( \mathbf{z} \sim \mathcal{U}(0, 1)^{256} \) is sampled from a uniform distribution. The generator maps this vector to a complex-valued quantum state, and the fidelity with the target state is evaluated. The computed loss is then backpropagated through the custom autograd function.
To stabilize training, all gradients are manually scaled by a constant \texttt{scaling\_factor} before each optimizer step. The training loop is executed for a fixed number of epochs (depending on the number of qubits), and the best fidelity observed across all epochs is recorded as a key performance indicator.

\subsubsection{\textbf{Final Evaluation}}
After training, the final state is compared classically to the target state via the inner product $|\bra{\psi}\ket{\phi_u}|^2$ for validation.
A fidelity value close to $1$ indicates successful reconstruction of the target state.

\subsection{Gradient Free Method: Evolutionary Strategy}\label{grad_free}

\begin{algorithm}\label{algo:QESwap}
\fontsize{8.0pt}{9.0pt}\selectfont
\caption{Gradient-free State Vector Estimation}
\begin{algorithmic}[1]
\State Initialize population size \( N \), noise scale \( \sigma \), learning rate \( \alpha \), and parameters \( \mathbf{w} \sim \mathcal{N}(0, I) \)
\For{each iteration up to max\_iter}
    \State Sample noise vectors \( \mathbf{z}_i \sim \mathcal{N}(0, I) \) for \( i = 1, \dots, N \)
    \State Generate candidates: \( \mathbf{w}_i = \mathbf{w} + \sigma \mathbf{z}_i \)
    \State Normalize \( \mathbf{w}_i \) and convert to \( |\psi_{\text{gen}}^{(i)}\rangle \)
    \State Evaluate fidelity \( F_i = \text{SWAPTest}( |\psi_{\text{gen}}^{(i)}\rangle, |\psi_{\text{unknown}}\rangle ) \)
    \State Compute advantages: \( A_i = \frac{F_i - \bar{F}}{\text{std}(F)} \)
    \State Update: \( \mathbf{w} \leftarrow \mathbf{w} + \frac{\alpha}{N \sigma} \sum_i A_i \mathbf{z}_i \)
    \If{fidelity exceeds threshold}
        \State \textbf{break}
    \EndIf
\EndFor
\end{algorithmic}
\label{alg:statevector_ES}
\end{algorithm}

While gradient-based methods achieved fidelities approaching 1 in noiseless simulations, they required extensive manual tuning of both neural network hyperparameters and the optimization strategy. This issue became more severe in the presence of noise, where gradient signals were often unstable, making training less reliable and substantially increasing the number of epochs required for convergence. For higher-dimensional quantum states, the achievable fidelity typically plateaued around 0.85, falling short of the desired 0.99.
A significant contributor to the overall computational cost was the gradient estimation process. Since the SWAP test is inherently non-differentiable, we employed finite difference methods to approximate gradients with respect to each component of the generated state vector. This approach proved to be computationally intensive, as it necessitated a large number of quantum circuit executions per optimization step.
To address these limitations, we explored gradient-free optimization techniques such as Evolutionary Strategies (ES) \cite{salimans_evolution_2017}.

We introduce \texttt{QESwap} \textit{(Quantum Evolutionary Strategy with SWAP Fidelity)}, a customized evolutionary strategy for reconstructing unknown quantum states using fidelity scores from the SWAP test. Unlike traditional gradient-based approaches, QESwap operates in a derivative-free regime, leveraging Gaussian perturbations and population-based search to iteratively maximize fidelity between a generated and unknown target state. Crucially, QESwap is designed for single-copy, non-destructive quantum learning, making it ideal for mid-circuit state reconstruction and dynamic memory applications. To our knowledge, this is the first evolutionary strategy framework to integrate SWAP-test-based fidelity estimation into a hybrid quantum-classical optimization loop.
Our method is implemented using \texttt{PennyLane} and \texttt{PyTorch}, integrating both classical and quantum components to iteratively optimize the representation of a target quantum state, be it a state vector, density matrix, or unitary matrix. 
Without loss of generality, we describe our approach using the state vector representation in Algorithm~\ref{alg:statevector_ES}.

\subsubsection{\textbf{Problem Formulation}}
Let the unknown quantum state be denoted by \( |\phi_{\text{unknown}}\rangle \), a normalized complex-valued vector in a \( 2^n \)-dimensional Hilbert space for \( n \) qubits. The goal is to estimate a parametrized vector \( |\psi_{\text{gen}}(\mathbf{w})\rangle \) such that the fidelity 
\(
F = \left| \langle \phi_{\text{u}} \mid \psi_{\text{gen}}(\mathbf{w}) \rangle \right|^2
\)
is maximized. This optimization is performed using our proposed QESwap algorithm, with the swap test implemented as a quantum circuit to estimate fidelity. The SWAP test as shown in section \ref{swap_test} is exactly the same as described in the gradient-based approach. 

\subsubsection{\textbf{Evolutionary Strategy Optimization}}

The ES algorithm searches for the optimal parameters \( \mathbf{w} \in \mathbb{R}^{2 \cdot 2^n} \), representing the real and imaginary parts of a candidate state. Each candidate state is parameterized as a vector \( \mathbf{w} \in \mathbb{R}^{2d} \), where \( d = 2^n \), with real and imaginary components flattened into a single vector. No assumptions are made about the structure or sparsity of the target state. The methodology of section \ref{algo:QESwap} proceeds as follows:
The methodology described in Section~\ref{algo:QESwap} begins by initializing the parameter vector \( \mathbf{w} \) using a Gaussian distribution. In each iteration, a population of \( N = 50 \) perturbed candidates is generated by adding Gaussian noise. Each candidate vector \( \mathbf{w} \in \mathbb{R}^{2d} \) is interpreted as a complex vector by pairing consecutive entries as real and imaginary components, and is normalized to unit norm to ensure physical validity before fidelity evaluation. The swap test fidelity is then computed for each candidate, and standardized rewards are used to update \( \mathbf{w} \) via the rule:
\(
\mathbf{w} \leftarrow \mathbf{w} + \frac{\alpha}{N \sigma} \sum_i A_i \cdot \mathbf{z}_i,
\)
where \( \alpha = 0.05 \) is the learning rate, \( \sigma = 0.1 \) is the noise scale, \( A_i \) is the normalized advantage of the \( i \)-th candidate, and \( \mathbf{z}_i \) is its noise vector.
Training is terminated when either the maximum iteration count (100) is reached or when fidelity exceeds predefined thresholds (e.g., 0.95 or 0.99). All generated candidates are normalized before fidelity evaluation.

\subsubsection{\textbf{Experimental Protocol}}
We conduct experiments for upto 6-qubit systems. For each configuration, 100 independent trials are performed with randomly sampled unknown quantum states. The QESwap algorithm is used to reconstruct each state, and the fidelity and convergence epoch for thresholds \( F \geq 0.95 \) and \(\geq 0.99 \) are recorded.
All simulations are performed on noisy and noiseless quantum simulators and only single qubit results are performed in hardware as mentioned in Table \ref{tab:all_results}.

\subsection{Snapshot at multiple points}
Our framework enables quantum state snapshots (one at a time) at arbitrary circuit intersections. States are re-prepared in each iteration. This allows snapshots at any point in the circuit, making multi-point introspection feasible on current hardware through repeated SWAP tests.


\subsection{Depositing and Withdrawing the Quantum State}

Once the target unknown states are retrieved, the corresponding complex-valued vectors are stored in classical memory using standard techniques. When queried, either through classical access methods or via QRAM, the stored states are reloaded into quantum circuits using state preparation routines.

\section{Results and Analysis}

We have tested our methodology in real quantum hardware, noisy simulations using IBM hardware noise models, and noiseless simulations using AerSimulator in Pennylane.
The details of the noise model are mentioned in Appendix \ref{noise_model}.

\textbf{Real Hardware Results:}
We have tested our methodology in IBM Superconducting qubit hardware (\texttt{ibm\_sherbrooke)}.
Due to a lack of resources on real hardware, we have verified only a few known single-qubit states, such as the states $\ket{0} = [1, 0]$, $\ket{1} = [0, 1]$, and the Hadamard state $\frac{1}{\sqrt{2}}[1, 1]$. We achieved a fidelity of $1$ within $3$ epochs. In these experiments, we have used the QESwap strategy as discussed in Section \ref{grad_free}, as it performed best in noisy simulations (explained later in this section).

\textbf{Entanglement Structure of Target and Reconstructed States:}
First, we prepared 100 random unknown states $\ket{\phi_u}$ (as mentioned in Section \ref{unknown_state_prep}) that were used in all the experiments unless otherwise mentioned.
To determine whether our unknown target states are sufficiently entangled, we measured the entanglement entropy of those states using the following method. We calculated the von Neumann entropy of the reduced density matrix obtained by performing a bipartition of the system. Specifically, for a given pure state $\lvert\psi\rangle$, we traced out half of the system to obtain the reduced density matrix $\rho_A = \mathrm{Tr}_B(\lvert\psi\rangle\langle\psi\rvert)$, and then computed the entanglement entropy as
\(
S(\rho_A) = -\mathrm{Tr}(\rho_A \log \rho_A).
\)
This entropy quantifies how entangled a state is across the chosen bipartition.
\begin{figure}
    \centering
    \includegraphics[width=1\linewidth]{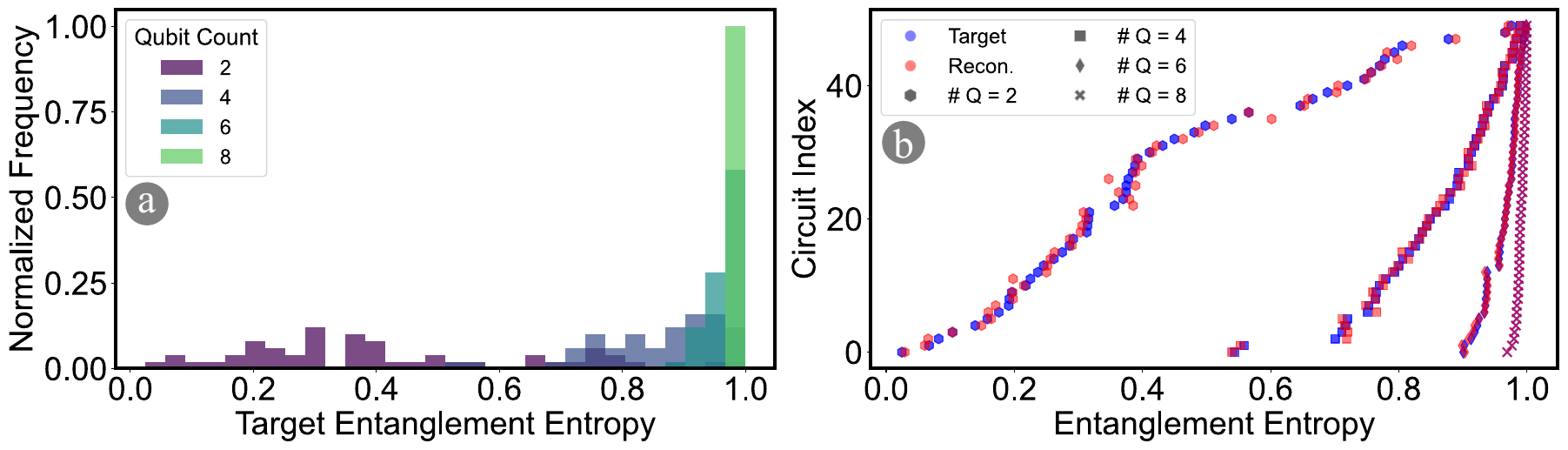}
    \caption{\textbf{Entanglement entropy analysis of target and reconstructed states:} (a) Distribution of target entanglement entropy across qubit counts. (b) Comparison of target and reconstructed entanglement entropy, sorted by circuit index, showing close alignment across different system sizes.}
    \label{fig:statevector_entanglement_entropy}
\end{figure}
We observed a good mixture of highly entangled states and matrix product states with low entanglement. Notably, the reconstructed states exhibited entanglement properties closely matching those of the corresponding unknown target states. This confirms the robustness of our method across a range of entangled and weakly entangled (matrix product) states. Fig.~\ref{fig:statevector_entanglement_entropy} \circled{a} displays the entanglement entropy of the target states, while Fig.~\ref{fig:statevector_entanglement_entropy} \circled{b} demonstrates that the entanglement entropy of the generated states aligns well with that of the target states. This strong agreement further supports the fact that the generated states faithfully capture the entanglement structure of the target states.

\textbf{Fidelity Distribution Analysis:}
\begin{figure}
    \centering
    \includegraphics[width=1\linewidth]{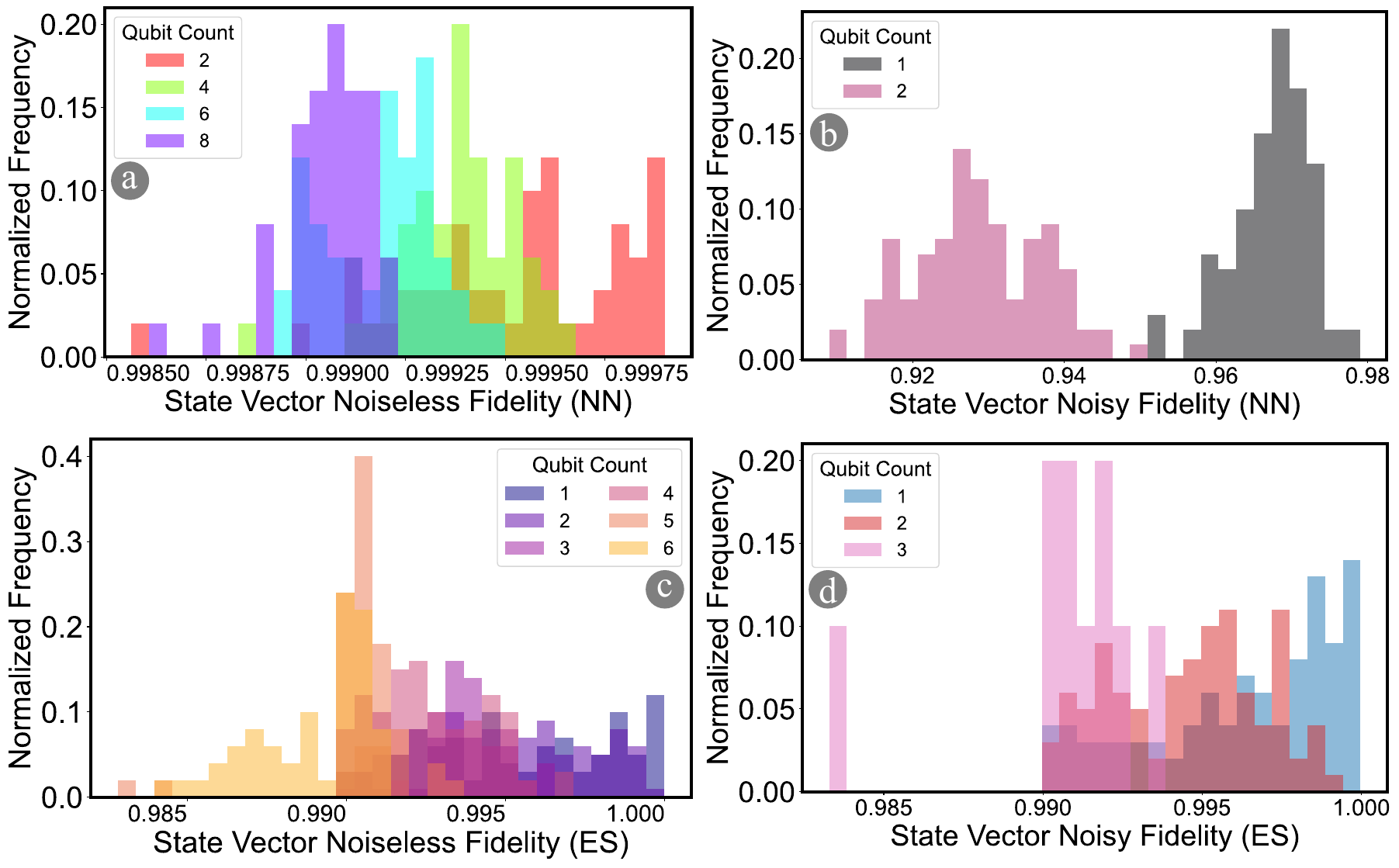}
    \caption{\textbf{Fidelity distributions for state vector reconstruction across qubit counts:} (a) \& (b) show results for the gradient-based (NN) method under noiseless and noisy settings; (c) \& (d) show the same for the gradient-free (ES) method.}
    \label{fig:statevector_fidelity_distribution}
\end{figure}
Fig. ~\ref{fig:statevector_fidelity_distribution} presents the fidelity values as defined in equation~\ref{eq:fidelity}. We evaluated fidelity by randomly sampling 1000 quantum states and reporting the corresponding fidelity for each reconstruction. 
Fig. ~\ref{fig:statevector_fidelity_distribution} \circled{a} and \circled{c} illustrate the results for noiseless simulations using the Neural Network (NN) and Evolutionary Strategy (ES) approaches, respectively. In both cases, fidelities approached values close to 1, indicating high accuracy in reconstructing the quantum states. Fig. ~\ref{fig:statevector_fidelity_distribution} \circled{b} and \circled{d} depict results from noisy simulations for NN and ES. We observed that learning under noise conditions proved to be significantly more challenging due to gradient instability. 

\begin{figure}
    \centering
    \includegraphics[width=1\linewidth]{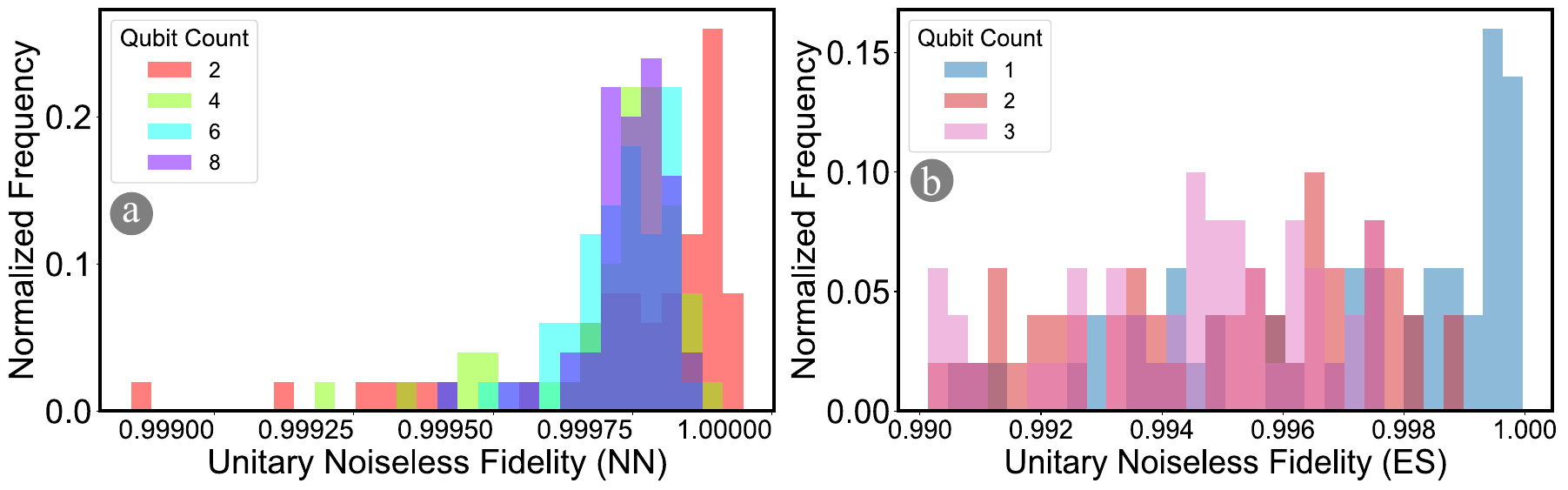}
    \caption{\textbf{Fidelity distributions for unitary matrix reconstruction} under noiseless conditions. (a) Gradient-based (NN) method; (b) Gradient-free (ES) method.}
    \label{fig:unitary_fidelity_distribution}
\end{figure}

Due to the increased simulation time required for noisy environments, we limited our evaluations: the NN-based reconstruction was tested up to 2 qubits, while the ES method was extended to 3 qubits. Interestingly, the ES method demonstrated faster convergence and was comparatively easier to configure, as shown in Table~\ref{tab:all_results}.
In the case of unitary matrix reconstruction, fidelities remained near 1 for noiseless simulations, as shown in Fig.~\ref{fig:unitary_fidelity_distribution} \circled{a} and \circled{b}. However, fidelity degraded quickly under noise as the number of qubits increased (Table~\ref{tab:quantum_state_result}). Consequently, we did not pursue the complete simulation of 100 random states for the unitary reconstruction under noisy conditions.

\begin{figure}
    \centering
    \includegraphics[width=1\linewidth]{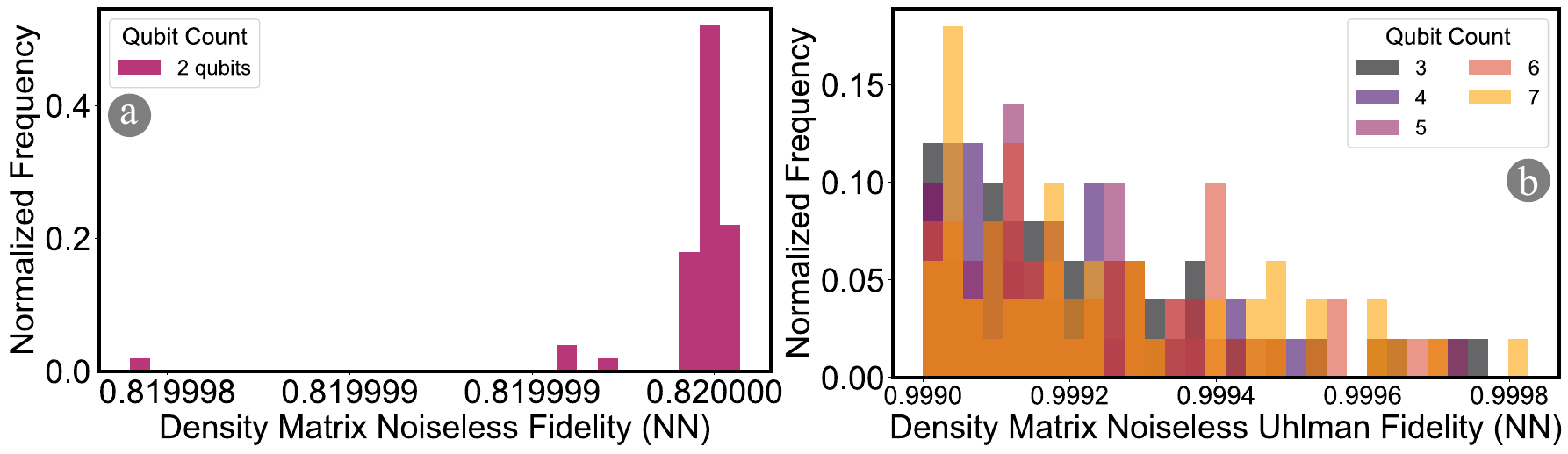}
    \caption{\textbf{Fidelity distributions for density matrix reconstruction} under noiseless conditions using the gradient-based (NN) method. (a) Trace fidelity for 2-qubit mixed states; (b) Uhlmann fidelity across larger number of qubits.}
    \label{fig:density_matrix_fidelity_distribution}
\end{figure}

For density matrix reconstruction using NN, even in the noiseless case with 2 qubits, the fidelity plateaued around 0.8 despite extensive hyperparameter tuning, as shown in Fig.~\ref{fig:density_matrix_fidelity_distribution} \circled{a}. This prompted us to investigate further. We discovered that the SWAP test fails to capture complete overlap information for mixed states, thus limiting its use for mixed-state fidelity estimation. Consequently, the exact reconstruction of mixed states via fidelity-based metrics becomes fundamentally infeasible.
To verify that our neural network and optimization strategies were not the limiting factors, we evaluated performance using the Uhlmann fidelity, which does reflect true overlap between mixed states, though it requires direct access to the unknown state's density matrix, which is impractical in real scenarios. Nevertheless, our method achieved near-perfect fidelity (as shown in Fig.~\ref{fig:density_matrix_fidelity_distribution} \circled{b}) using Uhlmann fidelity, confirming that our optimization techniques are effective. We therefore conclude that our method is well-suited for reconstructing pure quantum states, but not applicable to mixed states due to the inherent limitations of the SWAP test. As a result, the remainder of this paper will primarily focus on state vector and unitary regeneration techniques applied to pure states. Section~\ref{limitation_mixed} provides a deeper discussion on the nature and implications of this limitation, clarifying its practical impact.

\begin{table*}[t!]
\centering
\fontsize{7.0pt}{9.5pt}\selectfont
\caption{Average epochs to reach fidelity $> 0.99$ for 1, 2, and 3-qubit circuits across simulated and real backends computed over 100 random circuits and shown only for cases that successfully exceeded the threshold.}
\centering
\begin{tabular}{c|c||cccccc|cccccc|cc}
\hline \hline
\multirow{3}{*}{\textbf{\begin{tabular}[c]{@{}c@{}}Estimation\\ Strategies\end{tabular}}} & \multirow{3}{*}{\textbf{\begin{tabular}[c]{@{}c@{}}Regenerating\\ Methods\end{tabular}}} & \multicolumn{6}{c|}{\textbf{Noiseless Simulations}}                                                                                              & \multicolumn{6}{c|}{\textbf{Noisy Simulations}}                                                                                                  & \multicolumn{2}{c}{\textbf{Real Hardware}} \\ \cline{3-14}
                                                                                          &                                                                                          & \multicolumn{2}{c|}{\textbf{1 qubit}}              & \multicolumn{2}{c|}{\textbf{2 qubits}}             & \multicolumn{2}{c|}{\textbf{3 qubits}} & \multicolumn{2}{c|}{\textbf{1 qubit}}              & \multicolumn{2}{c|}{\textbf{2 qubits}}             & \multicolumn{2}{c|}{\textbf{3 qubits}} & \multicolumn{2}{c}{\textbf{1 qubit}}       \\ \cline{3-16} 
                                                                                          &                                                                                          & \textbf{\# E} & \multicolumn{1}{c|}{\textbf{Fid.}} & \textbf{\# E} & \multicolumn{1}{c|}{\textbf{Fid.}} & \textbf{\# E}      & \textbf{Fid.}     & \textbf{\# E} & \multicolumn{1}{c|}{\textbf{Fid.}} & \textbf{\# E} & \multicolumn{1}{c|}{\textbf{Fid.}} & \textbf{\# E}      & \textbf{Fid.}     & \textbf{\# E}        & \textbf{Fid.}       \\ \hline \hline
\multirow{2}{*}{\textbf{Gradient Based}}                                                  & \textbf{State Vector}                                                                    & 5             & \multicolumn{1}{c|}{0.9999}        & 12            & \multicolumn{1}{c|}{0.9997}        & 17                 & 0.9997            & 27            & \multicolumn{1}{c|}{0.9927}        & \multicolumn{4}{c|}{NA}                                                                     & \multicolumn{2}{c}{\multirow{2}{*}{NA}}    \\ \cline{2-14}
                                                                                          & \textbf{Unitary Matrix}                                                                  & 32            & \multicolumn{1}{c|}{0.9999}        & 48            & \multicolumn{1}{c|}{0.9999}        & 75                 & 0.9999            & \multicolumn{6}{c|}{NA}                                                                                                                          & \multicolumn{2}{c}{}                       \\ \hline
\multirow{2}{*}{\textbf{Gradient Free}}                                                   & \textbf{State Vector}                                                                    & 5             & \multicolumn{1}{c|}{0.9955}        & 7             & \multicolumn{1}{c|}{0.9951}        & 13                 & 0.9928            & 5             & \multicolumn{1}{c|}{0.9963}        & 8             & \multicolumn{1}{c|}{0.9945}        & 26                 & 0.9906            & 3                    & 0.99                \\ \cline{2-16} 
                                                                                          & \textbf{Unitary Matrix}                                                                  & 8             & \multicolumn{1}{c|}{0.9978}        & 11            & \multicolumn{1}{c|}{0.9949}        & 19                 & 0.9937            & 59            & \multicolumn{1}{c|}{0.99}          & \multicolumn{4}{c|}{NA}                                                                     & \multicolumn{2}{c}{NA}                     \\ \hline \hline
\end{tabular}
\label{tab:all_results}
\end{table*}
\begin{table*}
\centering
\caption{Maximum fidelity attained and epochs taken to reach it for selected standard 1, 2, and 3 qubit quantum states}
\fontsize{7.0pt}{9.5pt}\selectfont
\begin{tabular}{cc||cccccccc|cccccccc}
\hline \hline
\multicolumn{2}{c||}{\multirow{4}{*}{\textbf{Standard Quantum States}}}                                                                                                                                                                                                                               & \multicolumn{8}{c|}{\textbf{Noiseless Simulations}}                                                                                                                                                  & \multicolumn{8}{c}{\textbf{Noisy Simulations}}                                                                                                                                                      \\ \cline{3-18} 
\multicolumn{2}{c||}{}                                                                                                                                                                                                                                                                       & \multicolumn{4}{c|}{\textbf{Gradient Based (NN)}}                                                       & \multicolumn{4}{c|}{\textbf{Gradient Free (ES)}}                                           & \multicolumn{4}{c|}{\textbf{Gradient Based (NN)}}                                                       & \multicolumn{4}{c}{\textbf{Gradient Free (ES)}}                                           \\ \cline{3-18} 
\multicolumn{2}{c||}{}                                                                                                                                                                                                                                                                       & \multicolumn{2}{c|}{\textbf{State Vec.}}           & \multicolumn{2}{c|}{\textbf{Unitary}}              & \multicolumn{2}{c|}{\textbf{State Vec.}}           & \multicolumn{2}{c|}{\textbf{Unitary}} & \multicolumn{2}{c|}{\textbf{State Vec.}}           & \multicolumn{2}{c|}{\textbf{Unitary}}              & \multicolumn{2}{c|}{\textbf{State Vec.}}           & \multicolumn{2}{c}{\textbf{Unitary}} \\ \cline{3-18} 
\multicolumn{2}{c||}{}                                                                                                                                                                                                                                                                       & \textbf{\# E} & \multicolumn{1}{c|}{\textbf{Fid.}} & \textbf{\# E} & \multicolumn{1}{c|}{\textbf{Fid.}} & \textbf{\# E} & \multicolumn{1}{c|}{\textbf{Fid.}} & \textbf{\# E}     & \textbf{Fid.}     & \textbf{\# E} & \multicolumn{1}{c|}{\textbf{Fid.}} & \textbf{\# E} & \multicolumn{1}{c|}{\textbf{Fid.}} & \textbf{\# E} & \multicolumn{1}{c|}{\textbf{Fid.}} & \textbf{\# E}     & \textbf{Fid.}    \\ \hline \hline
\multicolumn{1}{c|}{\multirow{2}{*}{\textbf{1 Qubit}}}  & $\bm{|0\rangle}$, $\bm{|1\rangle}$                                                                                                                                                                                                & 4             & \multicolumn{1}{c|}{0.99}          & 32            & \multicolumn{1}{c|}{0.99}          & 4             & \multicolumn{1}{c|}{0.99}          & 8                 & 0.99              & 28            & \multicolumn{1}{c|}{0.99}          & 39            & \multicolumn{1}{c|}{0.74}          & 5             & \multicolumn{1}{c|}{0.99}          & 60                & 0.99             \\ \cline{2-2}
\multicolumn{1}{c|}{}                                   & $\bm{|+\rangle}$, $\bm{|-\rangle}$                                                                                                                                                                                                & 4             & \multicolumn{1}{c|}{0.99}          & 32            & \multicolumn{1}{c|}{0.99}          & 4             & \multicolumn{1}{c|}{0.99}          & 9                 & 0.99              & 29            & \multicolumn{1}{c|}{0.99}          & 41            & \multicolumn{1}{c|}{0.75}          & 5             & \multicolumn{1}{c|}{0.99}          & 61                & 0.99             \\ \hline
\multicolumn{1}{c|}{\multirow{2}{*}{\textbf{2 Qubits}}} & $\bm{|00\rangle}$, $\bm{|01\rangle}$, $\bm{|10\rangle}$, $\bm{|11\rangle}$                                                                                                                                                        & 10            & \multicolumn{1}{c|}{0.99}          & 48            & \multicolumn{1}{c|}{0.99}          & 6             & \multicolumn{1}{c|}{0.99}          & 10                & 0.99              & 75            & \multicolumn{1}{c|}{0.98}          & 125           & \multicolumn{1}{c|}{0.61}          & 8             & \multicolumn{1}{c|}{0.99}          & 108               & 0.40             \\ \cline{2-2}
\multicolumn{1}{c|}{}                                   & $\bm{|\Phi^+\rangle}$, $\bm{|\Phi^-\rangle}$, $\bm{|\Psi^+\rangle}$, $\bm{|\Psi^-\rangle}$                                                                                                                                        & 12            & \multicolumn{1}{c|}{0.99}          & 50            & \multicolumn{1}{c|}{0.99}          & 7             & \multicolumn{1}{c|}{0.99}          & 12                & 0.99              & 77            & \multicolumn{1}{c|}{0.97}          & 123           & \multicolumn{1}{c|}{0.60}          & 9             & \multicolumn{1}{c|}{0.99}          & 110               & 0.41             \\ \hline
\multicolumn{1}{c|}{\textbf{3 Qubits}}                  & \begin{tabular}[c]{@{}c@{}}$\bm{\text{GHZ}_0^+}$, $\bm{\text{GHZ}_0^-}$, $\bm{\text{GHZ}_1^+}$, $\bm{\text{GHZ}_1^-}$, \\ $\bm{\text{GHZ}_2^+}$, $\bm{\text{GHZ}_2^-}$, $\bm{\text{GHZ}_3^+}$, $\bm{\text{GHZ}_3^-}$\end{tabular} & 18            & \multicolumn{1}{c|}{0.99}          & 76            & \multicolumn{1}{c|}{0.99}          & 13            & \multicolumn{1}{c|}{0.99}          & 21                & 0.99              & 84            & \multicolumn{1}{c|}{0.92}          & 93            & \multicolumn{1}{c|}{0.34}          & 26            & \multicolumn{1}{c|}{0.99}          & 20                & 0.15             \\ \hline \hline
\end{tabular}
\label{tab:quantum_state_result}
\end{table*}

\textbf{Amplitude-Based Comparison of Reconstructed States:}
\begin{figure}
    \centering
    \includegraphics[width=0.8\linewidth]{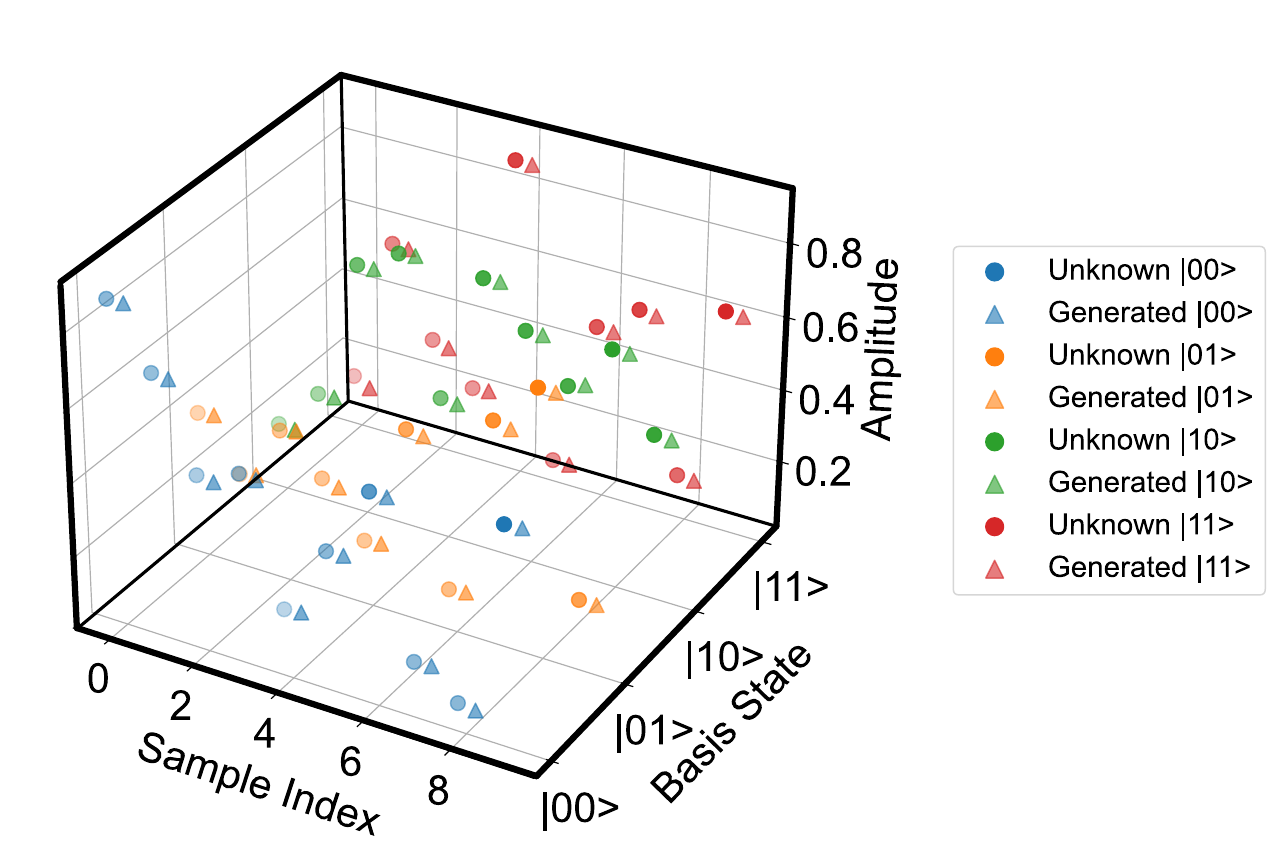}
    \caption{\textbf{Comparison of generated and target (unknown) quantum states} in the state vector representation. The amplitudes of each basis state ($\ket{00}$, $\ket{01}$, $\ket{10}$, $\ket{11}$) are shown across multiple samples. Close alignment of amplitudes indicates accurate reconstruction.}
    \label{fig:amplitude_3d_plot}
\end{figure}
We have picked a few specific quantum states to show in Fig. \ref{fig:amplitude_3d_plot} that amplitude probabilities in 2 qubit basis states are almost the same which reiterates the fact fidelity is high and also the right measure to compare similarity between quantum states.

\textbf{Fidelity Threshold Comparison:}
To evaluate the effectiveness and convergence behavior of our quantum state reconstruction models, we analyzed the number of epochs required to reach high-fidelity thresholds (specifically, fidelity $> 0.99$) across a variety of qubit configurations and simulation environments. Table~\ref{tab:all_results} summarizes the average number of epochs needed for different estimation strategies and regeneration methods in noiseless, noisy, and real hardware settings. If an epoch value is missing from the table, it indicates that the algorithm failed to converge to a fidelity of $0.99$.

In noiseless simulations, the gradient-based neural network (NN) model achieved fidelity $> 0.999$ within 5 to 17 epochs for state vector representations and required more epochs (32 to 75) for unitary matrix reconstruction. In contrast, the gradient-free evolutionary strategy (ES) consistently demonstrated faster convergence, requiring only 5 to 13 epochs for state vectors and fewer than 20 for unitaries.
Under noisy simulations, NN models showed slower convergence and increased variance in epoch counts, whereas ES remained robust, achieving $> 0.99$ fidelity in under 30 epochs for up to 3-qubit state vectors. On real quantum hardware, the ES approach reconstructed single-qubit states in just 3 epochs with fidelity reaching 0.99, demonstrating its practical viability and efficiency in resource-constrained environments.

\begin{figure}
    \centering
    \includegraphics[width=1\linewidth]{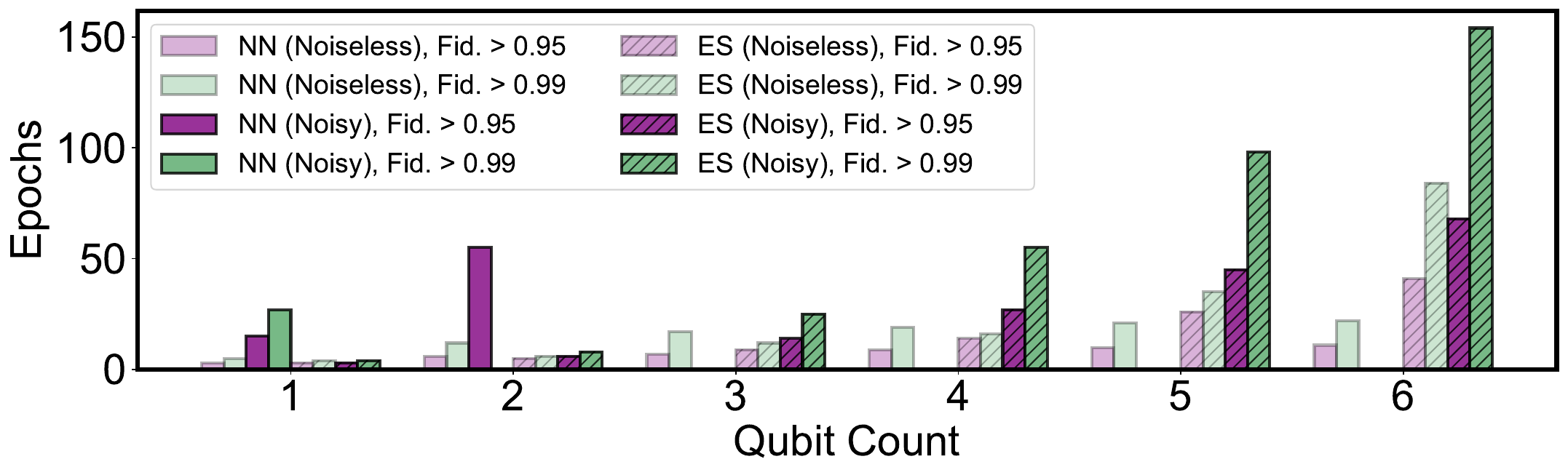}
    \caption{\textbf{Epochs required to reach fidelity thresholds (\bm{$>0.95$} and \bm{$>0.99$})} for state vector reconstruction across different qubit counts, under both noiseless and noisy conditions. Results are shown for gradient-based (NN) and gradient-free (ES) estimation strategies. The gradient-free method demonstrates faster and more consistent convergence, particularly in noisy settings.}
    \label{fig:grouped_barplot_correct_order}
\end{figure}

As observed in Table~\ref{tab:all_results}, not all combinations of regeneration methods and estimation strategies achieve a fidelity of $0.99$ under noisy simulations. Therefore, we restrict our analysis to the regeneration method based on state vector representation. Specifically, we examine the number of epochs required to reach fidelities of $0.95$ and $0.99$ using both gradient-based (NN) and gradient-free (ES) estimation strategies. The corresponding results are presented in Fig.~\ref{fig:grouped_barplot_correct_order}.
We observe that convergence to a fidelity of $0.95$ is generally more probable and requires fewer epochs compared to $0.99$. However, the gradient-based (NN) approach continues to struggle in reaching $0.95$ in certain noisy scenarios, whereas the gradient-free (ES) strategy consistently achieves both fidelity thresholds across noisy and noiseless simulations.
These findings combined, indicate that the gradient-free (ES) method provides superior stability and generalization in noisy and hardware-limited settings.

\textbf{Benchmarking on Standard Quantum States:}
To take our experiments one step further, we benchmarked our approach on a set of standard quantum states spanning 1 qubit basis states, 2 qubit Bell states, and 3 qubit GHZ states. Table~\ref{tab:quantum_state_result} summarizes the maximum fidelities attained and the number of epochs required to reach them, across both noiseless and noisy simulation settings. The comparison includes both Gradient-Based and Gradient-Free strategies and evaluates both state vector and unitary representations. Under noiseless conditions, both NN and ES achieve near-perfect fidelities ($\sim 0.99$) across all state types. However, NN methods consistently require fewer epochs than ES, especially for higher qubit systems. In noisy simulations, fidelity drops as expected. The impact is more significant for NN-based methods, while ES still achieves 0.99 (state vector) though it requires many more epochs. Therefore, gradient-free methods demonstrate greater robustness to noise, particularly in multi-qubit scenarios.

\textbf{Impact of Noise on Convergence Stability:}
It is prominent from Table~\ref{tab:all_results} and Table~\ref{tab:quantum_state_result} that noise significantly degrades performance, both in terms of fidelity and the number of epochs required to reach convergence. While in noiseless simulations, fidelities $\geq 0.99$ are often achieved within 20 epochs, noisy environments require many more iterations or even fail to converge, especially for gradient-based (neural network) methods. This degradation arises because, under noise, the model is effectively trying to estimate a moving target. As illustrated in Fig.~\ref{fig:bloch_sphere}, the noiseless case (left) shows a clear progression of reconstructed states toward the target state, whereas the noisy case (right) exhibits scattered and unstable convergence due to continuous fluctuations in the target state itself, complicating the optimization process.

\textbf{Selecting the Optimal Optimization Strategy:}
Based on our experiments and supporting reasoning, we find that the gradient-free estimation strategy, when combined with the state vector regeneration method, offers the most consistent performance across noiseless simulations, noisy conditions, and real hardware. It consistently converges to a fidelity of $\geq 0.99$, making it the most effective and reliable approach for snapshot-based quantum state reconstruction.

\section{Discussion}

\subsection{Applications of Quantum Snapshot}
Our proposed framework
has several transformative applications across quantum computing paradigms. Most notably, it enables \textit{non-destructive state introspection and classical storage}, effectively bridging the gap between quantum and classical memory architectures. By accurately reconstructing unknown quantum states with high fidelity and preserving them as classical vectors, our method enables a form of \textit{quantum memory bank}, which is retrievable on demand and compatible with existing state preparation primitives. 
%
Crucially, all the use cases (listed below) of quantum snapshot benefit from the \textit{hardware-agnostic} nature of our method, making it suitable for deployment on today’s NISQ-era quantum systems and inherently scalable as quantum hardware continues to advance.
\textbf{(a)} \textit{QRAM:} By capturing and storing intermediate quantum states during circuit execution, snapshots allow for precise state saving and retrieval. This significantly enhances the flexibility and efficiency of QRAM systems, enabling applications to access historical quantum states, reuse them across computational branches, or feed them into downstream tasks.
\textbf{(b)} \textit{Quantum Circuit Debugging:} Engineers and researchers can pause computation non-destructively, extract the quantum state, and inspect it classically—greatly aiding circuit validation and optimization.
\textbf{(c)} \textit{Modular Quantum Program Design:} Complex quantum programs can be decomposed into reusable modules, where intermediate states are stored and reloaded, reducing the need for complete circuit re-execution.
\textbf{(d)} \textit{Quantum Control and Feedback:} By supporting mid-circuit state estimation and re-preparation, our method enables adaptive control strategies essential for quantum error correction, feedback loops, and real-time control.
\textbf{(e)} \textit{Quantum Repeaters and Networking:} Accurate reconstruction and re-preparation of quantum states enable teleportation-style strategies in quantum communication, with classical storage at repeater nodes.
\textbf{(f)} \textit{Efficient Simulation and Benchmarking:} In simulation, our approach facilitates benchmarking by enabling comparisons between noisy hardware outputs and idealized evolutions, through reproducible and storable state approximations.

\subsection{Limitations of Quantum Snapshot for Mixed States} \label{limitation_mixed}

While our proposed method performs well for pure quantum states 
its extension to mixed quantum states introduces fundamental limitations. These arise not only from practical challenges in implementation, but also from the theoretical properties of the SWAP test and its relationship to true quantum fidelity.

\subsubsection{Implicit Measurement of the Hilbert-Schmidt Inner Product}
The SWAP test, when applied to mixed states $\rho$ and $\sigma$, does not estimate the standard quantum fidelity, but rather computes the Hilbert-Schmidt inner product:
\(
\operatorname{Tr}(\rho \sigma)
\).
This quantity coincides with the pure state fidelity only when both states are pure. However, for general mixed states, it overestimates similarity and lacks operational meaning in tasks such as quantum state discrimination or channel verification. As a result, our model, trained to maximize $\operatorname{Tr}(\rho \sigma)$, may converge to a state that is mathematically close in the Hilbert-Schmidt sense, but operationally distant in terms of true fidelity or trace distance.

\subsubsection{Limited Discrimination Power}
For general mixed states, the SWAP test offers only sub-optimal distinguishability compared to optimal quantum measurements. The probability of success in distinguishing two non-orthogonal mixed states using the SWAP test is upper bounded:
\(
P_{\text{success}} \leq \frac{3}{4}
\).
In contrast, optimal Helstrom measurements can achieve:
\(
P_{\text{success}} = 1 - \frac{1}{2} \operatorname{Tr}|\rho - \sigma|
\).
This theoretical gap highlights a key limitation: our training signal saturates well before the ideal distinguishability threshold is reached.
While our method offers an efficient, scalable, and differentiable approach for learning unknown pure quantum states, it is not designed for reconstructing general mixed states. The core limitation lies in both the structure of the generator and the interpretability of the SWAP test’s output. Future extensions should incorporate richer representations of quantum states and alternative fidelity estimators that align more closely with operational notions of state similarity in quantum information theory. 
The limitation of our method to pure states rather than general mixed states does not significantly hinder its practical utility in near-term quantum computing. Most target states in quantum algorithms (e.g., superposition states, entangled Bell states, or GHZ states) are intentionally designed as pure states, and even in noisy intermediate-scale quantum (NISQ) systems, the ideal intermediate computational states remain pure in theory. While environmental noise inevitably introduces mixed-state characteristics, our method’s high-fidelity pure-state reconstructions still provide operationally meaningful approximations for debugging, introspection, and memory applications—tasks that prioritize identifying dominant state components over characterizing full decoherence. Furthermore, critical use cases like QRAM, modular circuit design, and quantum state reuse inherently assume pure states as their input/output, aligning with our method’s strengths. This focus on pure states mirrors the operational assumptions of most quantum algorithms, making the mixed-state limitation a theoretical edge case rather than a practical barrier for current applications.









\section{Conclusion}

In this work, we introduced a novel, hardware-agnostic framework for non-destructively capturing and storing quantum states i.e. quantum snapshots, using classical memory. By leveraging fidelity-based optimization via SWAP tests, our method circumvents the limitations of conventional tomography, enabling dynamic introspection and modular reuse of quantum states during circuit execution. Our evaluation across simulated and real hardware demonstrates high-fidelity reconstructions with minimal overhead, particularly with the gradient-free QESwap strategy, which proves robust under noise. While limitations persist for mixed states, our approach offers a practical, scalable pathway toward classical quantum memory, laying critical groundwork for future architectures in quantum debugging, control, and QRAM-enabled workflows.

\section*{Acknowledgment}
We thank Archisman Ghosh, our dear labmate and friend, for his invaluable insights in reviewing our manuscript and for his generous assistance with related tasks.
The work is supported in parts by the National Science Foundation (NSF) (CNS-2129675 and CCF-2210963) and gifts from Intel.


\bibliography{ref_zotero_qram,ref}
\bibliographystyle{ieeetr}

\appendices \label{appendix}
\section{Noise Model} \label{noise_model}

To realistically simulate the behavior of near-term noisy quantum processors, we incorporate a detailed gate-level noise model derived from an IBM \texttt{GenericBackendV2} for the required number of qubits. This model is constructed using Qiskit's \texttt{NoiseModel} interface and imported into PennyLane via a custom Aer-based quantum device. The objective is to simulate realistic noise conditions akin to those found on current IBMQ hardware. Table \ref{tab:noiseparams} lists the values of the noise parameters.

\textbf{Basis Gate Set and Noisy Instructions:}
The simulated device supports the following native basis gate set:
\(
\{\texttt{cx},\ \texttt{delay},\ \texttt{id},\ \texttt{measure},\ \texttt{reset},\ \texttt{rz},\ \texttt{sx},\ \texttt{x}\}
\)
Among these, the subset subject to explicit noise modeling includes:
\(
\{\texttt{id},\ \texttt{cx},\ \texttt{sx},\ \texttt{x},\ \texttt{measure}\}
\)
This configuration reflects real-device behavior where idle operations, entangling gates, and standard unitary rotations are affected by gate-dependent stochastic noise.

\textbf{Gate-Level Error Modeling:}
To more accurately capture fidelity degradation at the hardware level, our model incorporates three calibrated error channels. First, \textit{bit flip errors} are simulated using a Pauli-X channel with a flip probability of \( p = 2.003e-04 \), representing a 0.1\% chance of bit flips occurring during gate operations. Second, we include a \textit{depolarizing error model}, with an error probability of 0.0017 (0.17\%) for single-qubit gates and a higher error rate of 0.02 (2\%) for two-qubit gates such as CNOT. Lastly, we model \textit{thermal relaxation effects} using relaxation parameters \( T_1 \sim 272\,\mu s \) and \( T_2 \sim 188\,\mu s \), along with a readout length of \( 1216\,\text{ns} \). Together, these calibrated noise sources offer a realistic and hardware-aware framework for evaluating quantum circuit fidelity in noisy environments.
The depolarizing and thermal relaxation errors are composed to create a composite noise model for CNOT gates. Single-qubit noise channels are applied to all instances of \texttt{rz}, \texttt{sx}, and \texttt{x} gates, and the bit-flip error is applied independently to capture additional stochastic behavior.

\textbf{Asymmetric and Hardware-Aware Modeling:}
The noise model enables fine-grained control over stochastic error processes using Qiskit's \texttt{QubitChannel} abstraction, which simulates noise through Kraus operators. A key feature of this model is its ability to capture \textit{asymmetric noise} across different gate–qubit pairs, reflecting the non-uniform error landscape observed in real hardware. The model also supports \textit{variable Kraus complexity}, where certain two-qubit gates (e.g., \texttt{cx}) may be represented by up to 16 Kraus operators, while simpler single-qubit gates typically use 3 to 4 operators. This asymmetry and variability closely mirror the behavior of actual IBMQ devices, enabling the simulation to reflect anisotropic and layout-specific noise. For instance, the gate \texttt{cx}(0,1) may be modeled using a \texttt{QubitChannel} with 16 Kraus operators, whereas \texttt{cx}(0,2) may require only 9.


\begin{table}
\fontsize{7.5pt}{9.5pt}\selectfont
\centering
\caption{Realistic Parameters Used in the Noise Model}
\begin{tabular}{c||cc}
\textbf{Parameter} & \textbf{Value} & \textbf{Description} \\
\hline \hline
Bit flip probability (X) & 2.003e-04 & Pauli-X error per gate \\
Single-qubit depolarizing error & 1.701e-02 & For \texttt{rz}, \texttt{sx}, \texttt{x} \\
Two-qubit depolarizing error & 0.02 & For \texttt{cx} gates \\
$T_1$ & $272.21\,\mu s$ & Relaxation time \\
$T_2$ & $188.1\,\mu s$ & Dephasing time \\
Readout length & $1216\,ns$ & Duration time for measurement \\
Kraus operators (\texttt{cx}(0,1)) & 16 & High noise interaction \\
Kraus operators (\texttt{cx}(0,2)) & 9 & Lower or different topology \\
\hline \hline
\end{tabular}
\label{tab:noiseparams}
\end{table}

\textbf{Effect on Simulation:}
This detailed noise model was integrated into both the training and evaluation phases of our quantum state reconstruction framework. By emulating realistic gate-dependent and hardware-informed noise processes, we ensure that our fidelity metrics, particularly those derived from dot product overlap and SWAP test comparisons, accurately reflect the limitations and characteristics of real quantum hardware. This alignment is essential for validating the practical deployability of the reconstructed quantum states on NISQ-era processors.

\end{document}